\documentclass[twocolumn, prl, floatfix, superscriptaddress]{revtex4}
\pdfoutput=1	% needed for the arXiv to properly compile with pdflatex

\usepackage[tracking=smallcaps]{microtype}	% for an arXiv submission, set \pdfoutput=1 near the top so it uses pdflatex

\usepackage[utf8]{inputenc} % set input encoding (not needed with XeLaTeX)
\usepackage{amsmath,amsthm,amsfonts}
\usepackage{color}
\usepackage[usenames,dvipsnames]{xcolor}
\usepackage{url}
\definecolor{DarkGray}{rgb}{0.3,0.3,0.3}
\usepackage[colorlinks=true,breaklinks, linkcolor=DarkGray, citecolor=DarkGray, urlcolor=DarkGray]{hyperref}

\newcommand{\abs}[1]{\left| #1 \right|}

\newcommand{\ket}[1]{\left\vert{#1}\right\rangle}

\newcommand{\eqnref}[1]{\hyperref[#1]{{(\ref*{#1})}}}
\newcommand{\thmref}[1]{\hyperref[#1]{{Theorem~\ref*{#1}}}}
\newcommand{\lemref}[1]{\hyperref[#1]{{Lemma~\ref*{#1}}}}
\newcommand{\corref}[1]{\hyperref[#1]{{Corollary~\ref*{#1}}}}
\newcommand{\defref}[1]{\hyperref[#1]{{Definition~\ref*{#1}}}}
\newcommand{\secref}[1]{\hyperref[#1]{{Section~\ref*{#1}}}}
\newcommand{\figref}[1]{\hyperref[#1]{{Fig.~\ref*{#1}}}}
\newcommand{\tabref}[1]{\hyperref[#1]{{Table~\ref*{#1}}}}
\newcommand{\remref}[1]{\hyperref[#1]{{Remark~\ref*{#1}}}}
\newcommand{\appref}[1]{\hyperref[#1]{{Appendix~\ref*{#1}}}}
\newcommand{\claimref}[1]{\hyperref[#1]{{Claim~\ref*{#1}}}}
\newcommand{\propref}[1]{\hyperref[#1]{{Proposition~\ref*{#1}}}}
\newcommand{\exampleref}[1]{\hyperref[#1]{{Example~\ref*{#1}}}}
\newcommand{\conjref}[1]{\hyperref[#1]{{Conjecture~\ref*{#1}}}}

\usepackage{graphicx}

\newcommand{\logical}[1]{{\overline{#1}}}
\newcommand{\CCZ}{\text{CCZ}}

\begin{document}

\title{Universal fault-tolerant quantum computation \\ with only transversal gates and error correction}
\author{Adam Paetznick}
\affiliation{David R. Cheriton School of Computer Science and Institute for Quantum Computing, University of Waterloo, Waterloo, Ontario N2L 3G1, Canada}
\author{Ben W. Reichardt}
\affiliation{Ming Hsieh Department of Electrical Engineering, University of Southern California, Los Angeles, California 90089, USA}
\date{\today}
         
\begin{abstract}
Transversal implementations of encoded unitary gates are highly desirable for fault-tolerant quantum computation.  Though transversal gates alone cannot be computationally universal, they can be combined with specially distilled resource states in order to achieve universality.
We show that ``triorthogonal'' stabilizer codes, introduced for state distillation by Bravyi and Haah [Phys. Rev. A \textbf{86} 052329 (2012)], admit transversal implementation of the controlled-controlled-$Z$ gate.
%Bravyi and Haah (2012) have proposed a state distillation procedure based on ``triorthogonal" stabilizer codes.  We observe that such codes also admit transversal implementation of the controlled-controlled-$Z$ gate, which is equivalent to the Toffoli gate up to single-qubit Hadamard rotations.  
We then construct a universal set of fault-tolerant gates \emph{without} state distillation by using only transversal controlled-controlled-$Z$, transversal Hadamard, and fault-tolerant error correction.  
We also adapt the distillation procedure of Bravyi and Haah to Toffoli gates, improving on existing Toffoli distillation schemes.
\end{abstract}

\maketitle

Quantum computers are highly susceptible to noise.  For protection, the data can be stored in a code~\cite{Shor1995a,Gottesman1997a}, and encoded operations can be applied ``fault-tolerantly," in order to prevent a single error from spreading to multiple qubits in a codeword~\cite{Shor1997}.  Between operations, fault-tolerant error-correction keeps errors from accumulating.  The simplest fault-tolerant operation is the application of physical gates transversally across the codewords, meaning that the $j$th gate is applied to the $j$th qubits of the codewords, for every~$j$.  Depending on the gate, this may or may not preserve the codespace and implement a valid encoded operation.  Unfortunately, no quantum code admits transversal implementation of a universal set of encoded gates~\cite{Eastin2009a}.  Instead, universality is usually achieved by combining some transversal gates with specially prepared resource states in a process known as state injection and distillation~\cite{Bravyi2004}.  State distillation can be orders of magnitude more costly than direct transversal gates, and dominates the resource overhead for implementing a quantum computer~\cite{Raussendorf2007,Fowler2013}.  

%\vspace{-.07cm}

Here we propose a way of implementing a universal set of quantum gates transversally, % for certain quantum error-correcting codes, 
up to a correction that can be made by the standard error-correction procedure.  In effect, our protocol shows that the impossibility theorem of~\cite{Eastin2009a} can be circumvented without adding any new machinery.
Separate injection and distillation procedures are not required.  The construction works only for the class of ``triorthogonal" quantum stabilizer codes, introduced recently by Bravyi and Haah~\cite{Bravyi2012a}.  %Because this is a restricted class of codes, 
Therefore, implementing our construction directly may not reduce the overhead compared to using state distillation with more efficient codes that can tolerate higher noise rates.  However, based on our construction, we derive a state-distillation procedure that, with realistic error parameters, reduces the overhead compared to previous state-of-the-art state distillation methods~\cite{Eastin2012,Jones2012d}.  

Our construction is based on two main insights.  First, we observe that the controlled-controlled-$Z$ operation (defined by $\CCZ \ket{a,b,c} = (-1)^{abc} \ket{a,b,c}$ for bits $a,b,c$) can be implemented transversally for any triorthogonal quantum code.  Second, we show that the Hadamard $H = \frac{1}{\sqrt 2}\left(\begin{smallmatrix} 1 & 1 \\ 1 & -1\end{smallmatrix}\right)$ can be implemented by transversal $H$ gates followed by stabilizer measurements and Pauli $X$ corrections.  Together, $H$ and $\CCZ$ are universal for quantum computation~\cite{Shi02,Aharonov2003}.

As an example, consider the $[[15,7,3]]$ quantum Hamming code, which uses $15$ qubits to protect $7$ encoded qubits to distance $3$.  The codespace is the simultaneous $+1$ eigenspace of the eight operators 
\begin{center}
\begin{tabular}{c@{$\,\!\!$}c@{$\,\!\!$}c@{$\,\!\!$}c@{$\,\!\!$}c@{$\,\!\!$}c@{$\,\!\!$}c@{$\,\!\!$}c@{$\,\!\!$}c@{$\,\!\!$}c@{$\,\!\!$}c@{$\,\!\!$}c@{$\,\!\!$}c@{$\,\!\!$}c@{$\,\!\!$}c@{$,\quad$}c@{$\,\!$}c@{$\,\!$}c@{$\,\!$}c@{$\,\!$}c@{$\,\!$}c@{$\,\!$}c@{$\,\!$}c@{$\,\!$}c@{$\,\!$}c@{$\,\!$}c@{$\,\!$}c@{$\,\!$}c@{$\,\!$}c@{$\,\!$}c@{,}}
$I$&$I$&$I$&$I$&$I$&$I$&$I$&$X$&$X$&$X$&$X$&$X$&$X$&$X$&$X$ & $I$&$I$&$I$&$I$&$I$&$I$&$I$&$Z$&$Z$&$Z$&$Z$&$Z$&$Z$&$Z$&$Z$ \\
$I$&$I$&$I$&$X$&$X$&$X$&$X$&$I$&$I$&$I$&$I$&$X$&$X$&$X$&$X$ & $I$&$I$&$I$&$Z$&$Z$&$Z$&$Z$&$I$&$I$&$I$&$I$&$Z$&$Z$&$Z$&$Z$ \\
$I$&$X$&$X$&$I$&$I$&$X$&$X$&$I$&$I$&$X$&$X$&$I$&$I$&$X$&$X$ & $I$&$Z$&$Z$&$I$&$I$&$Z$&$Z$&$I$&$I$&$Z$&$Z$&$I$&$I$&$Z$&$Z$ \\
$X$&$I$&$X$&$I$&$X$&$I$&$X$&$I$&$X$&$I$&$X$&$I$&$X$&$I$&$X$ & $Z$&$I$&$Z$&$I$&$Z$&$I$&$Z$&$I$&$Z$&$I$&$Z$&$I$&$Z$&$I$&$Z$
\end{tabular}
\end{center}
each the tensor product of Pauli operators $I$, $X$ and~$Z$.  
Choose a basis for this codespace so the logical $X$ and $Z$ operators on the first encoded qubit are transversal $X$ and $Z$, respectively.  Provided that the other six ``gauge" qubits are prepared as encoded $\ket{0^6}$,
the $\CCZ$ operation is transversal.  Moreover, transversal $H$ gates preserve the codespace and apply a logical $H$ to the first encoded qubit.  
Transversal $H$ corrupts the gauge qubits, but they can be restored to $\ket{0^6}$ by measuring the six corresponding logical $Z$ operators and applying $X$ corrections as necessary.  Measurement of the gauge qubits' logical $Z$ operators can be performed fault-tolerantly using standard error-correction techniques~\cite{Shor1997,Steane1996}.

In addition to allowing for universal quantum computation directly, our result also permits more efficient state injection and distillation when non-triorthogonal codes are used for computation.  Traditional injection and distillation procedures are used for fault-tolerantly implementing the $T = \left(\begin{smallmatrix}1&0 \\ 0 & e^{i\pi/4}\end{smallmatrix}\right)$ gate.
Recently, Eastin~\cite{Eastin2012} and Jones~\cite{Jones2012d,Jones2013a} have shown that the cost of implementing a Toffoli gate can be reduced by distilling so-called Toffoli \emph{states} rather than using fault-tolerant $T$ gates.  We observe that the distillation procedure of~\cite{Bravyi2012a} for $T$ gates can be adapted to Toffoli states in order to improve on the procedures of Eastin and Jones.

\section{Stabilizer codes based on triorthogonal matrices}
%\medskip
Let us begin by specifying the construction of stabilizer codes based on triorthogonal matrices. For two binary vectors $f, g \in \{0,1\}^n$, let $f\cdot g \in \{0,1\}^n$ be their entry-wise product, and let $\abs f$ denote the Hamming weight of~$f$.  
Call an $m\times n$ binary matrix $G$, with rows $f_1,\ldots,f_m \in \{0,1\}^n$, ``triorthogonal" if
\begin{align*}
|f_i\cdot f_j| &= 0 \!\!\!\pmod 2 &\text{and}\quad |f_i\cdot f_j\cdot f_k| &= 0 \!\!\!\pmod 2
\end{align*}
for all pairs $(i,j)$ and triples $(i,j,k)$ of distinct indices.  

An $m \times n$ triorthogonal matrix $G$ can be used to construct an $n$-qubit, ``triorthogonal," stabilizer code as follows~\cite{Bravyi2012a}. For each even-weight row of $G$, add a stabilizer by mapping non-zero entries to $X$ operators, e.g., $(1,0,1) \mapsto X \otimes I \otimes X$. Add a stabilizer for each row of the orthogonal complement $G^\perp$ by similarly mapping non-zero entries to $Z$ operators. The logical $X$ and $Z$ operators are then given by mapping non-zero entries of the odd-weight rows of $G$ to $X$ and $Z$, respectively.

For example, fixing the six gauge qubits of the $15$-qubit Hamming code to $\ket{0^6}$ gives a $[[15,1,3]]$ triorthogonal code~\cite{Knill1996a}.
Bravyi and Haah have constructed a $[[49,1,5]]$ and a family of $[[3k+8, k, 2]]$ triorthogonal codes~\cite{Bravyi2012a}.

\section{Toffoli construction}
\begin{figure}
\centering
\includegraphics[height=1.5cm]{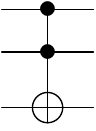}
\raisebox{.75cm}{~~\large{=}~~}
\includegraphics[height=1.5cm]{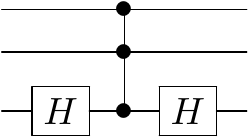}
\caption{\label{fig:toffoli-ccz-equivalence}
The Toffoli gate is equivalent to a $\CCZ$ gate in which the target qubit is conjugated by Hadamard gates.
}
\end{figure}

%\medskip
We next construct a fault-tolerant Toffoli gate for a triorthogonal code. 
Shown in \figref{fig:toffoli-ccz-equivalence}, the 
Toffoli gate is equivalent to a $\CCZ$ in which the target qubit is conjugated by Hadamard gates. 

\def\fodd {f_{\star}} %{f_1} {f_{\text{odd}}

The main component of the construction is an implementation of an encoded $\CCZ$ gate. We claim that for any triorthogonal code, transversal application of $\CCZ$ gates realizes $\CCZ$ gates on the encoded qubits. For simplicity consider the case of a triorthogonal code with a single encoded qubit, i.e., based on a triorthogonal matrix $G$ with a single odd-weight row, $f_{\star}$. (The argument with multiple encoded qubits is fully analogous.) Let $\mathcal{G}_0 \subseteq \{0,1\}^n$ be the linear span of all the even-weight rows of $G$ and let $\mathcal{G}_1$ be the coset $\{\fodd + g : g\in \mathcal{G}_0\}$. Then the encoding of $\ket a$, for $a \in \{0,1\}$, is given by the uniform superposition over $\mathcal{G}_a$: $\ket{\logical{a}} = \frac{1}{\sqrt{\abs{\mathcal{G}_a}}} \sum_{g \in \mathcal{G}_a} \ket g$.

The action of transversal $\CCZ$ on an encoded basis state $\ket{\logical{a,b,c}}$, for $a, b, c \in \{0,1\}$, is therefore given by 
\begin{equation}
\begin{split}
\CCZ^{\otimes n} \ket{\logical{a,b,c}} 
 &=\sum_{g \in \mathcal{G}_a, h \in \mathcal{G}_b, i \in \mathcal{G}_c} \CCZ^{\otimes n} \ket{g,h,i} \\
 &= \sum_{g \in \mathcal{G}_a, h \in \mathcal{G}_b, i \in \mathcal{G}_c} (-1)^{|g\cdot h\cdot i|}\ket{g,h,i}
\enspace.
\end{split}
\label{eq:transversal-ccz-on-basis-state}
\end{equation}
Here $g \cdot h \cdot i$ can be expanded as $(a \fodd + g')\cdot (b \fodd + h')\cdot (c \fodd + i')
$,
where $g', h', i' \in \mathcal{G}_0$. Expanding further gives one term $abc (\fodd \cdot \fodd \cdot \fodd) = abc \fodd$, plus other triple product terms in which $\fodd$ appears at most twice. Since $G$ is triorthogonal, these other terms necessarily have even weight. The term $abc \fodd$ has odd weight if and only if $a = b = c = 1$. Substituting back into~\eqref{eq:transversal-ccz-on-basis-state}, as desired, 
\begin{equation}
\CCZ^{\otimes n} \ket{\logical{a,b,c}} = (-1)^{abc} \ket{\logical{a,b,c}} %= \logical{\CCZ} \ket{\logical{abc}}
\enspace.
\end{equation}

To complete the Toffoli construction, we also require a fault-tolerant implementation of the Hadamard~$H$. For $H$ to be transversal, the code must be self-dual, i.e., $\mathcal{G}_0 = G^\perp$. Unfortunately, no triorthogonal code is self-dual. Indeed, otherwise, since $\CCZ$ is transversal it would be possible obtain a transversal implementation of Toffoli and $H$ for the same code.  However, Toffoli and $H$ together are universal~\cite{Shi02,Aharonov2003}, and it is known that universality cannot be achieved by transversal gates alone~\cite{Eastin2009a} (see also~\cite{Zeng2007,Chen2008a}).

Nonetheless, compact and fault-tolerant implementations of logical $H$ are still possible. When transversal $H$ is performed on a triorthogonal code, the logical operators are transformed properly: logical $X$ maps to logical $Z$ and vice versa. A subset of the stabilizers is preserved: observe that $\mathcal{G}_0 \subset G^\perp$, and thus each element of $\mathcal{G}_0$ corresponds to both $X$ and $Z$ stabilizers, which transversal $H$ swaps.  Transversal $H$ does not preserve the $Z$ stabilizers corresponding to $G^\perp \setminus \mathcal{G}_0$, so these must be restored by measuring and correcting them.  In the $[[15,7,3]]$ example above, this involved measuring the six gauge qubits' logical $Z$ operators.

\begin{figure}
\centering
\raisebox{.75cm}{\includegraphics[height=3cm]{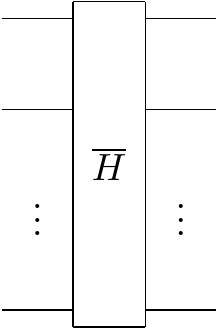}}
\raisebox{2.25cm}{~~\large{=}~~}
\includegraphics[height=4.5cm]{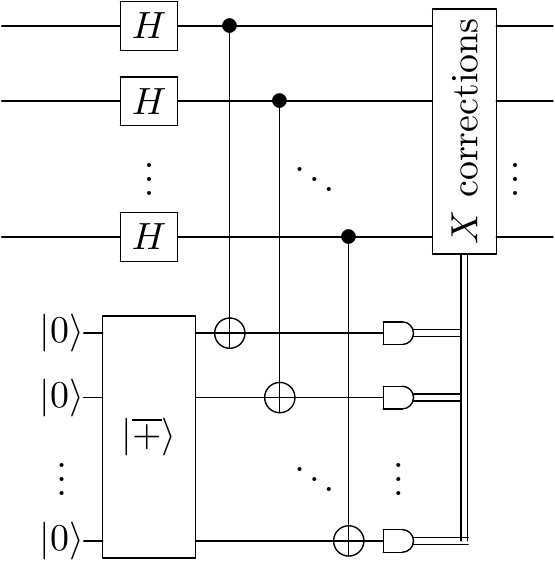}
\caption{\label{fig:steane-hadamard}
An implementation of the logical Hadamard operation in a triorthogonal code.  Transversal Hadamard gates are applied to the data block.  In order to restore the data to the codespace, and also correct any $X$ errors, an encoded $\ket +$ state is prepared, coupled to the data with transversal CNOT gates and measured. $X$ corrections are applied as necessary.
}
\end{figure}

The $Z$ stabilizers of $G^\perp \setminus \mathcal{G}_0$ can be restored during an $X$ error-correction procedure.  Steane's procedure, for example, involves a transversal CNOT from the data to an encoded $\ket + = \frac{1}{\sqrt 2}(\ket 0 + \ket 1)$ ``ancilla" state~\cite{Steane1996,Steane2004}.
%---implementing an encoded CNOT---followed by transversal $Z$-basis measurements of the ancilla.
The transversal CNOT implements encoded CNOT and has the effect of copying the unwanted $X$ stabilizers onto the ancilla, and copying the desired $Z$ stabilizers from the ancilla to the data.  Transversal $Z$-basis measurements of the ancilla then permit correcting $X$ errors on the data, while simultaneously restoring the stabilizer group.
(Once again, in the $[[15,7,3]]$ example, the ancilla's gauge qubits are prepared in $\ket{0^6}$. Each is measured as $0$ or $1$, with a correction required in the latter case.)  See~\figref{fig:steane-hadamard}.  
Alternatively, Shor's error-correction procedure uses a separate GHZ state $\frac{1}{\sqrt{2}}(\ket{00\ldots0} + \ket{11\ldots1})$ for each stabilizer to be measured~\cite{Shor1997}. 
In either case, the required stabilizers can be measured and corrected using $H$, $X$, CNOT, $\ket 0$ preparation and $Z$-basis measurements.  By using $\CCZ$ gates to simulate CNOT and $X$, the entire Toffoli requires only $H$ and $\CCZ$ gates.  

Importantly, even with additional $X$ corrections to fix the $Z$ stabilizers of $G^\perp \setminus \mathcal{G}_0$, the $X$ error-correction procedure is fault tolerant. That is, $k$ gate failures can lead to a data error of weight at most~$k$, for $k$ less than half the code's distance~$d$. Indeed, $d$ is the minimum of the code's distance $d_Z$ against $Z$ errors (determined by the $X$ stabilizers of $\mathcal{G}_0$) and its distance $d_X$ against $X$ errors (determined by the $Z$ stabilizers of $G^\perp$); and since $\mathcal{G}_0 \subset G^\perp$, $d_Z \leq d_X$. Thus any $X$ error of weight less than $d/2$ can be corrected using only the $Z$ stabilizers of $\mathcal{G}_0$, and the $Z$ stabilizers of $G^\perp \setminus \mathcal{G}_0$ can be corrected separately.  
Indeed, the Hadamard construction of~\figref{fig:steane-hadamard} works for any CSS code in which the $X$ and $Z$ logical operators have identical supports and transversal Hadamard conjugates the $X$ stabilizers to a subset of the $Z$ stabilizers.

% A similar construction works for a triorthogonal code that encodes more than one qubit, i.e., when the triorthogonal matrix $G$ has multiple odd-weight rows. The triorthogonality condition again implies that the phase induced by transversal $\CCZ$ depends only on the cosets to which each of the encoded qubits belong.  Symmetry of the logical $X$ and $Z$ operators permits a Hadamard implementation of the form given by~\figref{fig:steane-hadamard}.

\section{Discussion}
%\medskip
The simplest way to use the Toffoli construction above is with a concatenated triorthogonal code.  A universal set of fault-tolerant operations can be constructed from only $\CCZ$ and $H$ gates.  Thus using triorthogonal codes for computation could be useful for circuits that contain large numbers of Toffoli gates.  One could also imagine using multiple codes for computation by, for example, teleporting into the code best suited for each logical operation.  In this setting, a triorthogonal code could be used to implement efficiently the $\CCZ$ operation.  

Our construction allows for quantum computation to arbitrary accuracy, so long as the error rate per physical gate is below a constant ``threshold'' value~\cite{Aharonov1996a,Aliferis2005}.
Threshold error rates for triorthogonal codes are largely unknown, though estimates for the $[[15,1,3]]$ code are roughly $0.01$ percent per gate~\cite{Cross2009}.  If the $\CCZ$ operation is constructed from a sequence of one- and two-qubit gates, then the threshold is likely lower.
% Toffoli- and $\CCZ$-type gates have been demonstrated in a number of experimental settings,  with fidelities ranging from $68$ to $98$ percent~\cite{Monz2009,Mariantoni2011,Fedorov2012,Reed2012}. 
Since resource overhead increases rapidly as the physical noise rate approaches threshold, our construction is likely to be outperformed by schemes based on other codes, for which the threshold can be nearly one percent or higher~(see, e.g., \cite{Knill2004,Raussendorf2007,Wang2011}). The existence of high-performing triorthogonal codes is not out of the question, however.

A more conventional way to achieve universality is through state injection and distillation.  Bravyi and Haah have proposed distillation procedures using triorthogonal codes that permit fault-tolerant implementation of the $T$ gate~\cite{Bravyi2012a}.  Our result implies that a similar procedure could be used to implement Toffoli gates.

\begin{figure}
\raisebox{.75cm}{\includegraphics[height=2cm]{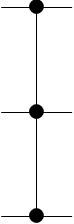}}
\raisebox{1.5cm}{~\large{=}~}
\includegraphics[height=3.5cm]{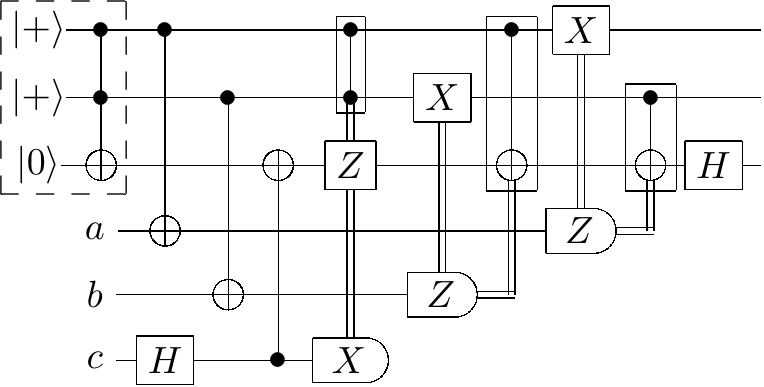}
\caption{\label{fig:ccz-gate-teleportation}
A $\CCZ$ gate can be implemented by consuming a single Toffoli state~\cite{Niel00}.  The input qubits are teleported into the Toffoli state (enclosed by the dashed line) with Clifford corrections conditioned on the measurement outcomes.}
\end{figure}

\begin{figure}
\includegraphics{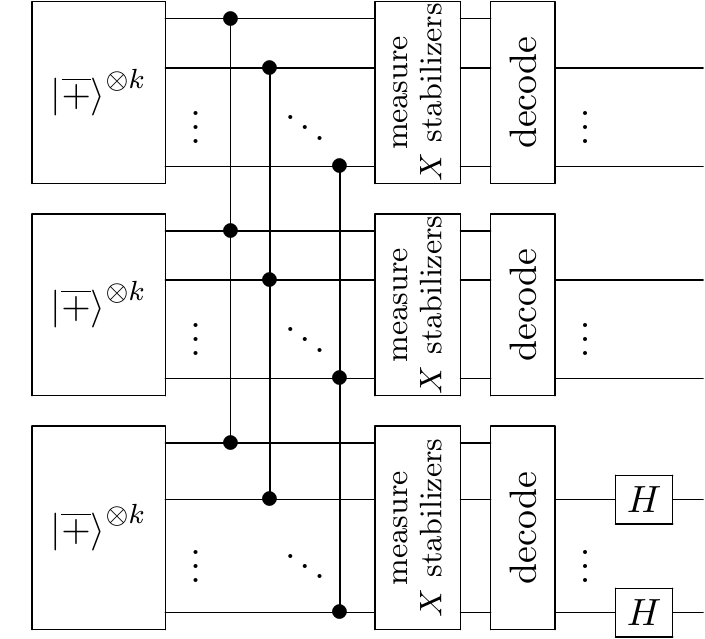}
\caption{\label{fig:toffoli-distillation}
A Toffoli state distillation circuit using a triorthogonal code encoding $k$ qubits.  Three separate blocks are encoded into the state $\ket{+}^{\otimes k}$ and then transversal $\CCZ$ gates are applied.  Conditioned on detecting no errors, each block is decoded and Hadamard gates are applied to each of the target qubits, yielding $k$ Toffoli states.}
\end{figure}

The Toffoli state is defined by the output of the Toffoli gate on input $\ket{+,+,0}$, where the third qubit is the target.  A Toffoli state can be used to implement the $\CCZ$ gate as shown in~\figref{fig:ccz-gate-teleportation}.
Distillation with a $[[3k+8,k,2]]$ triorthogonal code uses $3k+8$ transversal $\CCZ$ gates with error rate $p$ to produce $k$ Toffoli states with error rate $(3k+1)p^2$, to leading order in $p$. See~\figref{fig:toffoli-distillation}.  The $\CCZ$ gates in the distillation circuit can be implemented recursively, or by using an existing Toffoli distillation procedure~\cite{Jones2012d,Eastin2012,Jones2013a}.  Note that the Hadamard gates are performed {after} decoding and thus the circuit in~\figref{fig:steane-hadamard} is not required.  For simplicity, Clifford operations including encoding and decoding operations are assumed to be perfect.

The distillation procedure given by~\figref{fig:toffoli-distillation} can reduce the cost of implementing a Toffoli gate.
Suppose we wish to implement a Toffoli gate with error below $10^{-13}$.  The procedure of~\cite{Jones2012d} consumes eight $T$ gates with error $p$ to produce a Toffoli state with error $28p^2$.  The $T$ gates can be implemented using a combination of protocols; Table I of~\cite{Jones2012c} lists optimal protocol combinations for a large range of target error rates. If physical $T$ gates can be performed with error at most $10^{-2}$, then using the Toffoli construction of~\cite{Jones2012d}, as given, requires on average $540.16$ $T$ gates.  

\begin{figure}
\centering
\includegraphics[width=.48\textwidth]{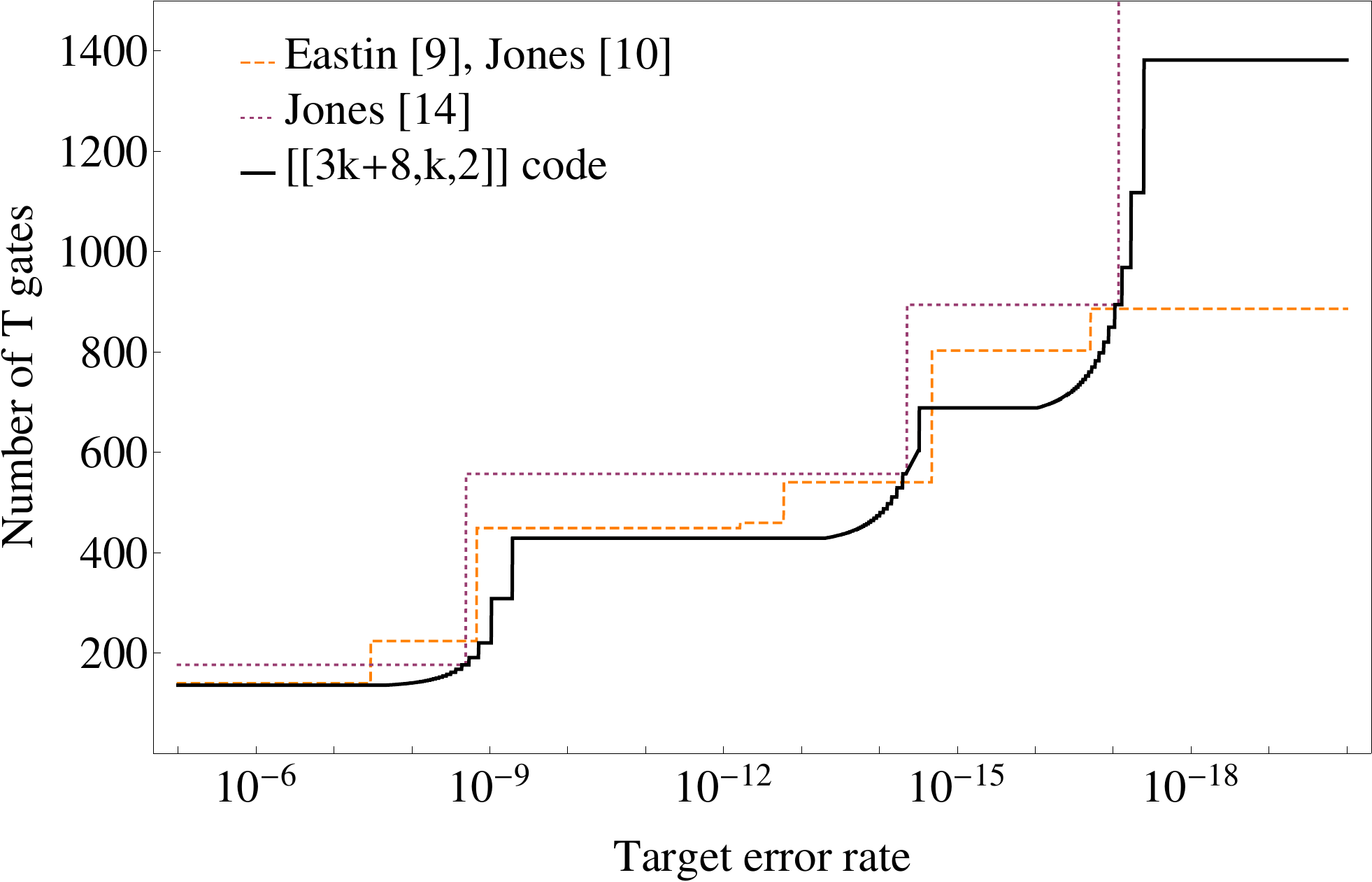}
\caption{\label{fig:t-cost-plot} 
The average number of physical $T$ gates required for three different Toffoli state distillation protocols.  For the previous protocols of~\cite{Eastin2012,Jones2012d} and~\cite{Jones2013a}, input $T$ gates are first distilled to the appropriate fidelity according to Table I of~\cite{Jones2012c}.  The solid black line shows the cost of our protocol for $[[3k+8,k,2]]$ triorthogonal codes where an even integer $2 \leq k \leq 100$ has been optimally selected at each target error rate. Input $\CCZ$ gates to the triorthogonal protocol are produced using~\cite{Jones2012d}. Physical $T$ gates are assumed to have error at most~$10^{-2}$.}
\end{figure}

Alternatively, we could use a $[[3k+8,k,2]]$ triorthogonal code for distillation at the top level, and use Toffoli states from~\cite{Jones2012d} as input to implement the $\CCZ$ gates in~\figref{fig:toffoli-distillation}. The distillation circuit fails to detect a faulty Toffoli state input only if the number of errors on each triorthogonal code block is even. To leading order, this occurs only if a pair of input Toffoli states contain identical errors.  Each of the seven possible errors on the output of states from~\cite{Jones2012d} are equally likely. Thus, if the input Toffoli states have error $p_1$, then to leading order the failure probability of the triorthogonal protocol is given by $7(3k+1)(p_1/7)^2$. For $k=100$, this yields an average $T$-gate cost of $428.7$, a savings of $25$\% over~\cite{Jones2012d} alone. Calculations for a range of target error rates are shown in~\figref{fig:t-cost-plot}.

The $T$ gate cost alone is an incomplete measure of the overhead.  Indeed,~\figref{fig:t-cost-plot} shows that the double error-detecting protocol of~\cite{Jones2013a} usually has higher $T$ gate cost than the single error-detecting protocol. However, the double error-detecting protocol can still yield significant savings since smaller code distances may be used for Clifford gates in intermediate distillation levels~\cite{Fowler2013,Jones2012d,Jones2013a}.  Our protocol similarly allows for reduced Clifford gate costs and offers the flexibility to be used recursively or on top of any other Toffoli state distillation protocol, including~\cite{Eastin2012,Jones2012d} and~\cite{Jones2013a}.  Complete overhead calculations depend on architectural considerations.  %architecture choices that are beyond the scope of this article.

Finally, we note that if the orthogonality conditions on the matrix $G$ are increased, then additional types of diagonal operations are transversal.  If $G$ satisfies the condition that all $j$-tuple products have weight $(0\mod 2)$ for all $2 \leq j \leq h$, then the $h$-fold controlled-$Z$ gate is transversal in the corresponding stabilizer code.
This observation is similar to a result of Landahl and Cesare, who demonstrated that codes satisfying increasingly stringent conditions on weights of the codewords admit transversal $Z$-axis rotations of increasing powers of $1/2^k$~\cite{Landahl2013}.

%\medskip
\section*{Acknowledgements}
The authors would like to thank Cody Jones for helpful feedback.
Supported in part by the Intelligence Advanced Research Projects Activity
(IARPA) via Department of Interior National Business Center contract number D11PC20166. The U.S. Government is authorized to reproduce and distribute reprints
for Governmental purposes notwithstanding any copyright annotation thereon. Disclaimer: The views and conclusions contained herein are those of the authors and
should not be interpreted as necessarily representing the official policies or endorsements, either expressed or implied, of IARPA, DoI/NBC, or the U.S. Government.
\vspace{-.5cm}

%\bibliographystyle{apsrev4-1}
%\bibliography{library}
%

\end{document}